\documentstyle[12pt,epsf]{article}
\newcommand{\eq}[1]{(\ref{#1})}
\newcommand{\ab}{( \alpha - \beta )} 
\newcommand{\nn}{\nonumber}
\newcommand{\fr}{\frac}
\newcommand{\mh}{m_h^2}
\newcommand{\mH}{m_H^2}
\newcommand{\mg}{m_G^2}
\newcommand{\ma}{m_A^2}

\newcommand{\al}{\alpha}
\newcommand{\be}{\beta}
\newcommand{\sn}{\sin}
\newcommand{\cs}{\cos}
\newcommand{\hs}{\hspace{1cm}}
\newcommand{\lt}{\left}
\newcommand{\rt}{\right}

\begin{document}
\begin{flushright}
  OU-HET 233 \\
\end{flushright}
\begin{center}
  {\bf SCREENING THEOREM FOR THE HIGGS DECAYS 
                                 INTO THE GAUGE BOSON PAIRS }
\footnote{Talk presented by Shinya Kanemura, 
           E-mail: kanemu@phys.wani.osaka-u.ac.jp}
 \end{center}
\vspace{0.5cm}
\begin{center}
   SHINYA KANEMURA, TAKAHIRO KUBOTA  and  HIDE-AKI TOHYAMA 
\end{center}
\begin{center}
     {\it Department of Physics, Osaka University \\
          Toyonaka, Osaka 560, Japan }  \\
    \end{center}
\begin{center}
  ABSTRACT 
\begin{quotation}
 The radiative corrections to the decays of the 
 neutral CP-even Higgs boson $H$ into 
 a longitudinal gauge boson pair, {\it i.e.}, 
 $H \rightarrow Z_LZ_L$ and $W^+_LW^-_L$ are analyzed
 in the two Higgs doublet 
 model by making use of the equivalence theorem. 
 The sensitivity of the decay rates to the masses of the heavier Higgs 
 bosons, charged $G^{\pm}$ 
 and CP-odd neutral $A$ bosons as well as CP-even neutral $h$ boson,  
 is investigated. 
 Though the width $\Gamma (H \rightarrow Z_LZ_L)$ 
 is insensitive to the masses of heavier Higgs bosons, 
 $\Gamma (H \rightarrow W^+_LW^-_L)$ is sensitive 
 and the radiative corrections are minimized for  $m_G = m_A$. 
 These results are explained completely on the basis of   
 a new screening theorem for the vertices, which is closely connected with
 the custodial $SU(2)_V$ symmetry.   
\end{quotation}
\end{center}

\section{Introduction} 

\hspace{12pt}
The discovery of the Higgs boson is the most important task for 
future accelerators including the next linear colliders. 
Let us suppose that the Higgs boson $H$ would be discovered
in these accelerators and that the mass would turn out to be above 
the threshold of a gauge boson pair.   
The Higgs sector is then expected to be a strong-coupled sector.   
The decay of the $H$ boson is then dominated by the decay processes 
into the longitudinal gauge boson pair, 
$H \rightarrow W^+_LW^-_L$ and $Z_LZ_L$,  
and these decay widths would be investigated experimentally to a 
considerable extent.  
Since studies on the non-decoupling effects due to internal particles in 
the radiative corrections are expected to be helpful to the approach 
to the strong Higgs sector, 
we have calculated the radiative corrections to these decay rates
in the two Higgs doublet model (THDM)
and have investigated the virtual effects on these due to 
the internal heavy scalar bosons, charged $G^{\pm}$, 
CP-odd neutral $A$ and the another CP-even neutral $h$ boson.
These results are summarized in ({\bf I}) and ({\bf II}) below. 
\begin{description}
\item[({I})]  
The radiative corrections to the decay width 
$\Gamma (H \rightarrow W^+_LW^-_L)$
are sensitive to the masses of $G^{\pm}$ and $A$ bosons ($m_G$ and $m_A$, 
respectively)   
but insensitive to the mass of $h$ boson 
and are minimized if we set $m_G = m_A$. 
\item[({II})]
The radiative corrections to the width 
$\Gamma (H \rightarrow Z_LZ_L)$ are relatively small and 
are insensitive to 
the masses of all the internal heavy scalar bosons.   
\end{description}

The main purpose of this talk is rather 
to present a new screening theorem: 
\begin{description}
\item[Theorem:]In the radiative corrections to 
these decay widths the leading contributions with respect to  
the masses of the internal heavy scalar bosons ($G^{\pm}$, $A$ and $h$)  
cancel out 
 in the custodial $SU(2)_V$ symmetric limit of the model.
\end{description}
The characteristic features of the decay widths mentioned in ({\bf I}) 
and ({\bf II}) can be given satisfactory explanations in terms  of this 
theorem.
This theorem reminds us of the Veltman's screening theorem \cite{vel}: 
the leading contributions of the Higgs boson mass  
in the oblique-type corrections cancel out by virtue of the 
custodial $SU(2)_V$ symmetry 
which becomes exact in the weak $U(1)_Y$-coupling limit in the standard model 
with one Higgs doublet.  
Veltman's theorem 
has been proved by Einhorn and Wudka \cite{ew} to all orders. 
In view of the similarity between Veltman's theorem     
for oblique-type corrections and our counterpart for vertices, 
our theorem is likely to hold 
to all orders of the perturbation, though we confirm this theorem 
only at one loop level.   

In our calculation,  
since we assume that  
$H$ boson is the lightest of all the Higgs bosons 
and is much heavier than the gauge bosons,  
we will make good use of the equivalence theorem. 
The calculations in the one Higgs-doublet model from the same viewpoint 
as ours are seen in ref.\cite{mw}. 
Other works related to the decays are given in ref.\cite{kni}.

\section{Two Higgs doublet model}

\hspace{12pt}
To begin with, we define the Higgs potential with two Higgs
doublets, $\Phi_1$ and $\Phi_2$. 
We would like to impose the discrete symmetry under 
$\Phi_2 \rightarrow - \Phi_2$ on the quartic-couplings in the potential  
 to avoid  in a natural way the flavor changing neutral current. 
Then the most general potential becomes 
\begin{eqnarray}
  V( \Phi_1 , \Phi_2 ) & = & - \mu_1^2 \left| \Phi_1 \right|^2
                             - \mu_2^2 \left| \Phi_2 \right|^2
                             - \lt( \mu_{12}^2 \Phi_1^{\dagger} \Phi_2 
                                     + {\rm \,h.c. \,} \rt)     \nn  \\
                       &   & + \lambda_1 \left| \Phi_1 \right|^4
                             + \lambda_2 \left| \Phi_2 \right|^4
                             + \lambda_3 \left| \Phi_1 \right|^2 
                                \left| \Phi_2 \right|^2 \nn \\
                       &   & + \lambda_4 \left( 
                                         {\rm Re }\Phi_1^{\dagger}
                                                  \Phi_2
                                                            \right)^2
                             + \lambda_5 \left(
                                         {\rm Im }\Phi_1^{\dagger}
                                                  \Phi_2
                                                            \right)^2 .
\label{pot}
\end{eqnarray}
Though the soft breaking term 
$- \lt( \mu_{12}^2 \Phi_1^{\dagger} \Phi_2 + \cdot\cdot\cdot \rt)$  
in eq.\eq{pot} becomes important in SUSY like models, we here set 
$\mu_{12}$ for zero 
because our interests are rather in the strong coupling situation. 
The effects of $\mu_{12}$ are then suppressed in the heavy mass limit.   
Therefore, the potential becomes to have seven parameters,  
$\mu_1$, $\mu_2$, $\lambda_1$, $\sim$, $\lambda_5$.

Note that the potential \eq{pot} would have
the custodial $SU(2)_V$ symmetry, which is the diagonal part of 
$SU(2)_L \otimes SU(2)_R$,  if $\lambda_5$ would be  zero. 
To see this, it is convenient to rewrite
eq.\eq{pot} in terms of $2 \times 2$ matrices 
${\bf \Phi}_i = \left(  i \tau_2 \Phi_i^{\ast}, \Phi_i \right)$. 
Then the $\lambda_5$-term in \eq{pot} becomes   
\begin{eqnarray}
  \lambda_5
   \left\{{\rm tr}(\tau_3\bf{\Phi}_1^{\dagger} \bf{\Phi}_2)\right\}^2,
\label{cus}
\end{eqnarray}
and we can easily see that the term \eq{cus} breaks 
$SU(2)_R$ and thus $SU(2)_V$ symmetry explicitly. 
On the other hand, all the other parts in eq.\eq{pot} can be rewritten 
as the combinations of 
${\rm tr}({\bf{\Phi}}_i^{\dagger}{\bf{\Phi}}_j)\;(i, j =1, 2)$, which 
are clearly custodial $SU(2)_V$ symmetric.

The field configurations in the Higgs doublets are parameterized as   
\begin{eqnarray}
 \Phi_i = \left( \begin{array}{c}
                w^+_i                                              \\
                 \frac{1}{\sqrt{2}} ( h_i +v_i + i z_i )
                   \end{array} 
                   \right) \,\,,\, ( i = 1,2 ), \nn
\end{eqnarray}   
where the vacuum expectation values 
$v_1$ and $v_2$ are combined to give  
 $v = \sqrt{v_1^2 + v_2^2} \sim 246$GeV.
The diagonalization of the mass terms is performed by introducing two
kinds of mixing angles $\al$ and $\be$ in the following way; 
\begin{eqnarray}
\left( \begin{array}{c}
              h_1                                  \\
              h_2 
        \end{array}  \right)
        =  \left( \begin{array}{cr}
                       \cos \alpha & -\sin \alpha  \\
                        \sin \alpha & \cos \alpha
                       \end{array}  \right)
               \left( \begin{array}{c}
                       h                           \\
                       H 
                        \end{array}  \right),&&       \nn      \\
\left( \begin{array}{c}
              w^{\pm}_1                                  \\
              w^{\pm}_2 
        \end{array}  \right)
        =   \left( \begin{array}{cr}
                       \cos \beta & -\sin \beta  \\
                        \sin \beta & \cos \beta
                       \end{array}  \right)
               \left( \begin{array}{c}
                       w^{\pm}                           \\
                       G^{\pm} 
                        \end{array}  \right),&&\;
\left( \begin{array}{c}
              z_1                                  \\
              z_2 
        \end{array}  \right)
        =   \left( \begin{array}{cr}
                       \cos \beta & -\sin \beta  \\
                        \sin \beta & \cos \beta
                       \end{array}  \right)
               \left( \begin{array}{c}
                       z                           \\
                       A 
                        \end{array}  \right) .      \nn  
\end{eqnarray}
We set $\tan \be = v_2 / v_1$ as usual, so that 
fields $w^{\pm}, z$ would be Nambu-Goldstone bosons. 
There are four massive fields, namely, $H$, $h$, $G^{\pm}$ and $A$. 

The five quartic-coupling constants in eq.\eq{pot} are expressed by the
masses of these scalar bosons together with the mixing angles;
\begin{eqnarray}
  \lambda_1 
&=& \fr{1}{2 v^2 \cs^2 \be} (\mh \cs^2 \al + \mH \sn^2 \al), \nn \\
  \lambda_2 
&=& \fr{1}{2 v^2 \sn^2 \be} (\mh \sn^2 \al + \mH \cs^2 \al), \nn \\
  \lambda_3 
&=& \fr{\sn 2\al}{v^2 \sn 2\be} (\mh - \mH) + \fr{2 \mg}{v^2}, \nn \\
  \lambda_4 
&=& - \fr{2 \mg}{v^2} , \nn \\
  \lambda_5 
&=& \fr{2}{v^2} (\ma - \mg).
\label{lam5}
\end{eqnarray}
Since only the $\lambda_5$-term breaks $SU(2)_V$, 
eq.\eq{lam5} means that 
the deviation from the degeneracy between $G^{\pm}$ and $A$ 
thus measures the explicit $SU(2)_V$ breaking. 
The seven independent parameters 
($\mu_1, \mu_2, \lambda_1, \sim, \lambda_5$) in eq.\eq{pot}  
are replaced by  
the four mass parameters ($m_h, m_H, m_G$ and $m_A$), 
two mixing angles ($\al$ and $\be$) and vacuum expectation value $v$.

\section{Radiative corrections}

\hspace{12pt}
Since we assume that all the Higgs masses are much greater than the gauge 
boson masses, we can make use of the equivalence theorem.  
The equivalence theorem at loop level is expressed as 
\begin{eqnarray}
&&  T \lt( Z_L(p_1), \cdot \cdot \cdot, Z_L(p_n),
         W_L(q_1), \cdot \cdot \cdot, W_L(q_m); \phi_a \rt) \nn \\ 
&& \;\;\; = 
   (C^Z_{\rm mod})^n(C^W_{\rm mod})^m
            T \lt(i z(p_1), \cdot \cdot \cdot, i z(p_n), 
                  i w(q_1), \cdot \cdot \cdot, i w(q_m); \phi_a \rt)
                        +  {\cal O} \lt( \fr{M_W}{\sqrt{s}} \rt), \nn
\end{eqnarray}
where 
$\phi_a$'s denote the other particles including Higgs bosons, 
$\sqrt{s}$ is the typical energy scale of the scattering process and 
$C^Z_{\rm mod}$ and $C^W_{\rm mod}$ are modification factors to be attached
to each external line of $Z_L$ and $W_L$'s respectively.   
Systematic studies of the general proof of the precise formulation of
the equivalence theorem at loop level 
have been presented by He, Kuang, and Li \cite{he}.
According to their work, the modification factors 
without fermion contributions turn out to be unity  
if we work in the Landau gauge,  on-mass-shell
renormalization scheme  and heavy-Higgs-mass limit.     
The important fact is that 
this statement is kept unchanged even if we work in THDM. 
The modification factors are, however, to receive additional contributions 
due to quark loops, which will be discussed in detail in ref.\cite{kkt}.  
Thus we have only to calculate the radiative corrections to the processes 
$H \rightarrow w^+ w^-$ and $ z z $ to evaluate those to the processes 
$H \rightarrow  W^+_LW^-_L$ and  $Z_L Z_L$.
Though the equivalence theorem does not mention
anything about the internal particles, we can neglect all the 
diagrams with internal $W^{\pm}$ and  $Z$ propagators because they are
suppressed by $M_W^2/m_H^2$. 
After all, our calculations are reduced simply to those in the
Higgs-Goldstone system with top quark.

Now let us prepare the counter-terms for the decay processes.
We start from the case of the process 
$H \rightarrow w^+w^-$.
The tree level interaction dictating this process is    
extracted from the potential \eq{pot} as 
\begin{eqnarray}
  {\cal L}_{Hww}= \fr{\mH}{v} \sn \ab 
          H w^+ w^- . \label{vtx} 
\end{eqnarray}
Some of the counter-terms required in the one loop calculation 
for this process is obtained by varying the parameters 
in eq.\eq{vtx}, namely, by putting 
$\mH \rightarrow \mH - \delta \mH , 
 v \rightarrow v - \delta v , 
\al \rightarrow \al - \delta \al$ , 
and 
$\be \rightarrow \be - \delta \be$.
Others 
come from the renormalizations of wave-functions and state-mixings  between 
those fields having the same quantum numbers, {\it i.e.},
by imposing the following replacement upon 
the bare interaction terms of $Hw^+G^-, HG^+w^-$ and $hw^+w^-$ 
as well as eq.\eq{vtx};
\begin{eqnarray}
\left( \begin{array}{c}
              h                                  \\
              H 
        \end{array}  \right)
        &\rightarrow&   \left( \begin{array}{cr}
                        \sqrt{Z_h}    &  \sqrt{Z_{hH}}  \\
                        \sqrt{Z_{Hh}} &  \sqrt{Z_H}
                                   \end{array}  \right)
                        \left( \begin{array}{cr}
                       1   &  - \delta \alpha  \\
                       \delta \alpha   &    1
                       \end{array}  \right)
               \left( \begin{array}{c}
                       h                           \\
                       H 
                        \end{array}  \right)   \nn    \\
\left( \begin{array}{c}
              w                                  \\
              G 
        \end{array}  \right)
        &\rightarrow&   \left( \begin{array}{cr}
                        \sqrt{Z_w}    &  \sqrt{Z_{wG}}  \\
                        \sqrt{Z_{Gw}} &  \sqrt{Z_G}
                                   \end{array}  \right)
                        \left( \begin{array}{cr}
                       1   &  - \delta \beta  \\
                       \delta \beta   &    1
                       \end{array}  \right)
               \left( \begin{array}{c}
                       w                           \\
                       G 
                        \end{array}  \right)     \nn \\
\left( \begin{array}{c}
              z                                  \\
              A
        \end{array}  \right)
        &\rightarrow&   \left( \begin{array}{cr}
                        \sqrt{Z_z}    &  \sqrt{Z_{zA}}  \\
                        \sqrt{Z_{Az}} &  \sqrt{Z_A}
                                   \end{array}  \right)
                        \left( \begin{array}{cr}
                       1   &  - \delta \beta  \\
                       \delta \beta   &    1
                       \end{array}  \right)
               \left( \begin{array}{c}
                       z                           \\
                       A
                        \end{array}  \right).    \nn  
\end{eqnarray}
After setting $\sqrt{Z_{hH}} = \sqrt{Z_{Hh}}$ and  
$\sqrt{Z_{wG}} = \sqrt{Z_{Gw}}$,  the full counter-term for 
the process $H \rightarrow w^+w^-$ is obtained as follows; 
\begin{eqnarray}
  \delta {\cal L}_{Hww} &=& 
      \lt[  \lt( - \fr{\delta \mH}{\mH} + \fr{\delta v}{v} \rt)
        \fr{\mH}{v} \sn \ab   
  - \fr{\mH}{v} (\delta \al - \delta \be) \cs \ab  \rt.\nn \\
  &+& \lt\{ \lt( \sqrt{Z_H} - 1 \rt) + \lt(Z_w - 1 \rt)\rt\}
      \fr{\mH}{v}\sn \ab  \nn \\
&-& \lt. \lt( \sqrt{Z_{hH}} - \delta \al\rt)\fr{\mh}{v}\cs \ab  
      - 2 \lt( \sqrt{Z_{wG}} + \delta \be \rt)\fr{\mH - \mg}{v}\cs \ab 
        \rt] Hw^+w^-. \nn
\end{eqnarray}
Similarly, setting $\sqrt{Z_{zA}} = \sqrt{Z_{Az}}$,
we also obtain the counter-term for the process, $H \rightarrow zz$;
\begin{eqnarray}
  \delta {\cal L}_{Hzz} &=& \lt[
          \lt( - \fr{\delta \mH}{\mH} + \fr{\delta v}{v} \rt)
        \fr{\mH}{2v} \sn \ab   
  - \fr{\mH}{2v} (\delta \al - \delta \be) \cs \ab  \rt. \nn \\
   && + \lt\{ \lt( \sqrt{Z_H} - 1 \rt) + \lt( Z_z - 1 \rt)\rt\}
      \fr{\mH}{2v}\sn \ab  \nn \\
&& \lt.- \lt( \sqrt{Z_{hH}} - \delta \al\rt)\fr{\mh}{2v}\cs \ab  
      -  \lt( \sqrt{Z_{zA}} + \delta \be \rt)\fr{\mH - \ma}{v}\cs \ab 
     \rt]
   Hzz. \nn
\end{eqnarray}

We are now full-fledged to perform the one-loop calculations of 
the amplitudes ${\cal M}_{Hww}(p^2)$ and ${\cal M}_{Hzz}(p^2)$ 
for the processes 
$H \rightarrow w^+w^-$ and $H \rightarrow zz$. 
The renormalization is performed in the  on-mass shell scheme. 
Here $\delta \be$ is defined by $zA$ mixing but not by $wG$ mixing. 
Details of the calculations will be explained in ref.\cite{kkt}.   
Finally we arrive at the decay width formula for each process,   
\begin{eqnarray}
\Gamma(H \rightarrow W^+_LW^-_L)
     &=& \fr{1}{16\pi} \fr{1}{m_H}
          \sqrt{1 - \fr{4M^2_W}{\mH}} 
          \left| {\cal M}_{Hww} (p^2 = \mH) \right|^2
          \left| C^W_{\rm mod} \right|^4, \label{Hww}\\   
\Gamma(H \rightarrow Z_LZ_L)
     &=& \fr{1}{32\pi} \fr{1}{m_H}
          \sqrt{1 - \fr{4M^2_Z}{\mH}} 
          \left| {\cal M}_{Hzz} (p^2 = \mH) \right|^2
          \left| C^Z_{\rm mod} \right|^4.\label{Hzz}
\end{eqnarray}

\section{Numerical analysis of the decay widths}

\hspace{12pt}
In this section, we would like to show some of our numerical 
analyses for the decay width formulae \eq{Hww} for 
$H \rightarrow W^+_LW^-_L$ (seen in part in ref.\cite{kk}), 
and  \eq{Hzz} for $H \rightarrow Z_LZ_L$. 
More details of our numerical results will be presented in ref.\cite{kkt}.

Some comments on the choice of the parameters are in order.
We choose the top-quark mass for 174 GeV and set 
the mass of the lightest Higgs boson $m_H$ tentatively for 300GeV 
throughout this talk. 
The mixing angle $\be$ is constrained to some 
extent by the low energy experimental data \cite{bb}, 
$\tan \be$ is not so smaller than unity.
 We therefore consider either of three cases, 
$\tan \be = 2, 10$ or $20$. 
The mixing angle $\al$ is less bounded phenomenologically \cite{gran} and so  
we vary  this parameter arbitrarily for theoretical interests.  
As to the masses of $h, G^{\pm}, A$ bosons, these are considerably 
constrained as a combination with $m_H$ and mixing angles from the 
analysis of the $\rho$ parameter \cite{rho} 
and especially $m_G$ is bounded as $m_G > 250$GeV from the data 
\cite{charged}. 
Here we set $m_h = 400$GeV  
and the masses of $G^{\pm}$ and $A$ bosons are varied as 
$300 < m_G < 900$GeV and $300 < m_A <1000$GeV. 
These parameter regions are all within the 
unitarity bounds \cite{kkta}. 
For the sake of the best illustration of the features 
({\bf I}) and ({\bf II}), 
we mainly show the case $\sin^2 (\al - \be) = 1$ below. 
The tree level evaluation for the decay widths are then calculated as
\begin{eqnarray}
  \Gamma_{\rm tree} (H \rightarrow W_L^+W^-_L) = 7.5{\rm GeV}, 
\;\,\,\Gamma_{\rm tree} (H \rightarrow Z_LZ_L) = 3.5{\rm GeV}.\label{tree}
\end{eqnarray}
Fig.1 shows the sensitivity of the width formulae
 \eq{Hww} and \eq{Hzz} to $m_G$ ($300 < m_G < 900$GeV)
 with $\tan \be = 2$ and with 
$m_A=$ 400, 700 and $1000$GeV 
(lines $a$, $b$ and $c$ for $\Gamma(H \rightarrow W^+_LW^-_L)$ 
and lines $a'$, $b'$ and $c'$ for $\Gamma (H \rightarrow Z_LZ_L)$ 
respectively). 
        
Looking at fig.1, we can see easily that $\Gamma (H \rightarrow Z_LZ_L)$ 
is quite insensitive to $m_G$ and $m_A$. 
On the other hand, we can also see   
that $\Gamma (H \rightarrow W^+_LW^-_L)$ is sensitive to $m_G$ and 
$m_A$ and that radiative corrections are minimized at $m_G = m_A$ 
(recall eqs.\eq{tree}). 
The difference of the behavior  
between the decay widths 
seen in fig.1 will be discussed in the next section.

Fig.2 shows the mixing angle $\al$ dependence of 
$\Gamma (H \rightarrow Z_LZ_L)$
for  $m_G= 500, m_A=600$GeV, and $\tan \be =2, 10$ and $20$.

\begin{center}      
\begin{minipage}[t]{7.8cm} 
\epsfxsize=7.8cm
\epsfbox{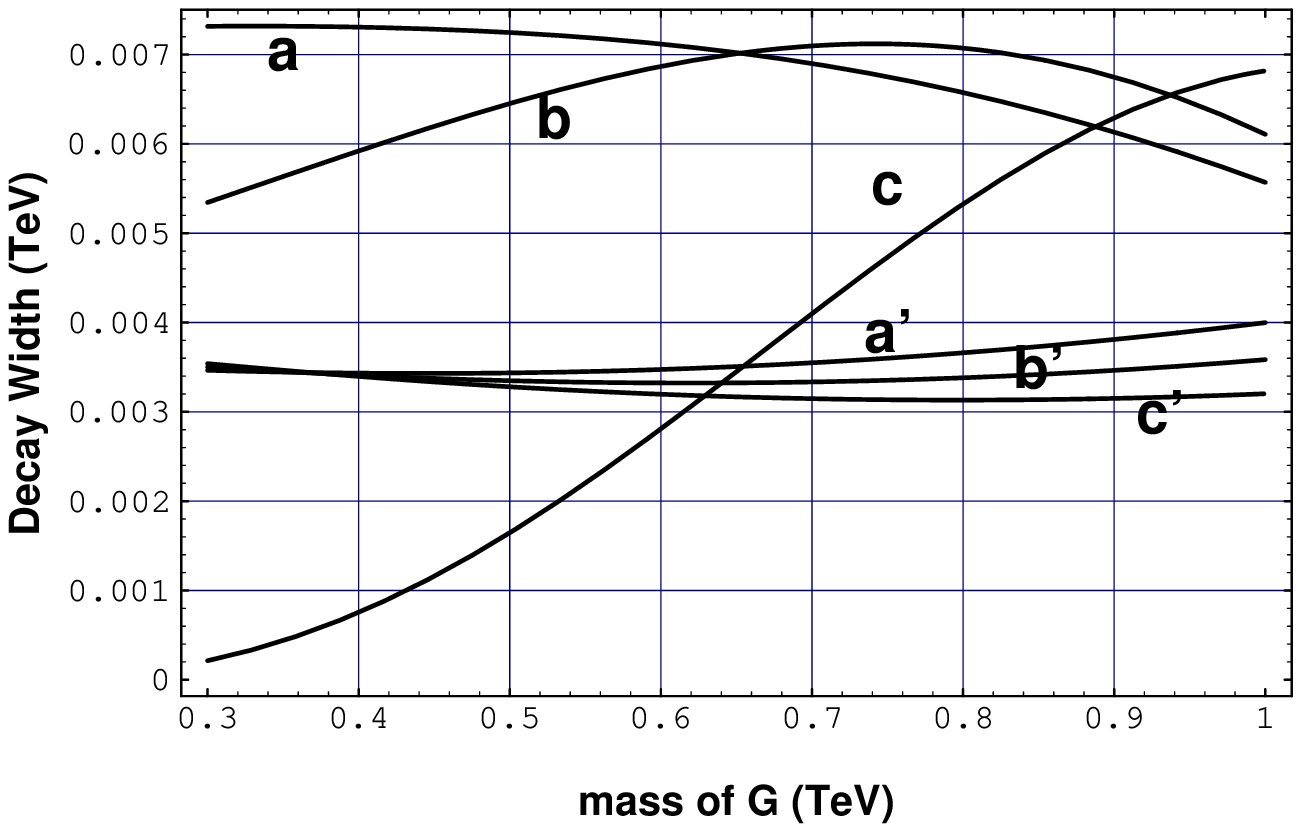}
\small
  {\bf Figure 1. } The decay widths as a function of $m_G$. 
The mixing angles are determined by $\tan \be = 2$ and 
$\sin^2 (\al - \be) = 1$. 
The masses of the neutral Higgs bosons are assumed to be 
$m_H = 300$ GeV and $m_h = 400$ GeV. 
Lines $a, b,$ and $c$ represent the behavior of 
$\Gamma (H \rightarrow W^+W^-)$
(eq. \eq{Hww}), while lines $a', b'$ and $c'$ represent that of   
 $\Gamma (H \rightarrow Z_LZ_L)$ (eq. \eq{Hzz}). 
The CP-odd Higgs boson mass is taken as $m_A = 400$ GeV ($a$ and $a'$), 
$m_A =700$ GeV ($b$ and $b'$) and $m_A = 1$ TeV ($c$ and $c'$), 
respectively. 
\normalsize
\end{minipage}
\hspace{1mm}
\begin{minipage}[t]{7.8cm} 
\epsfxsize=7.8cm
\epsfbox{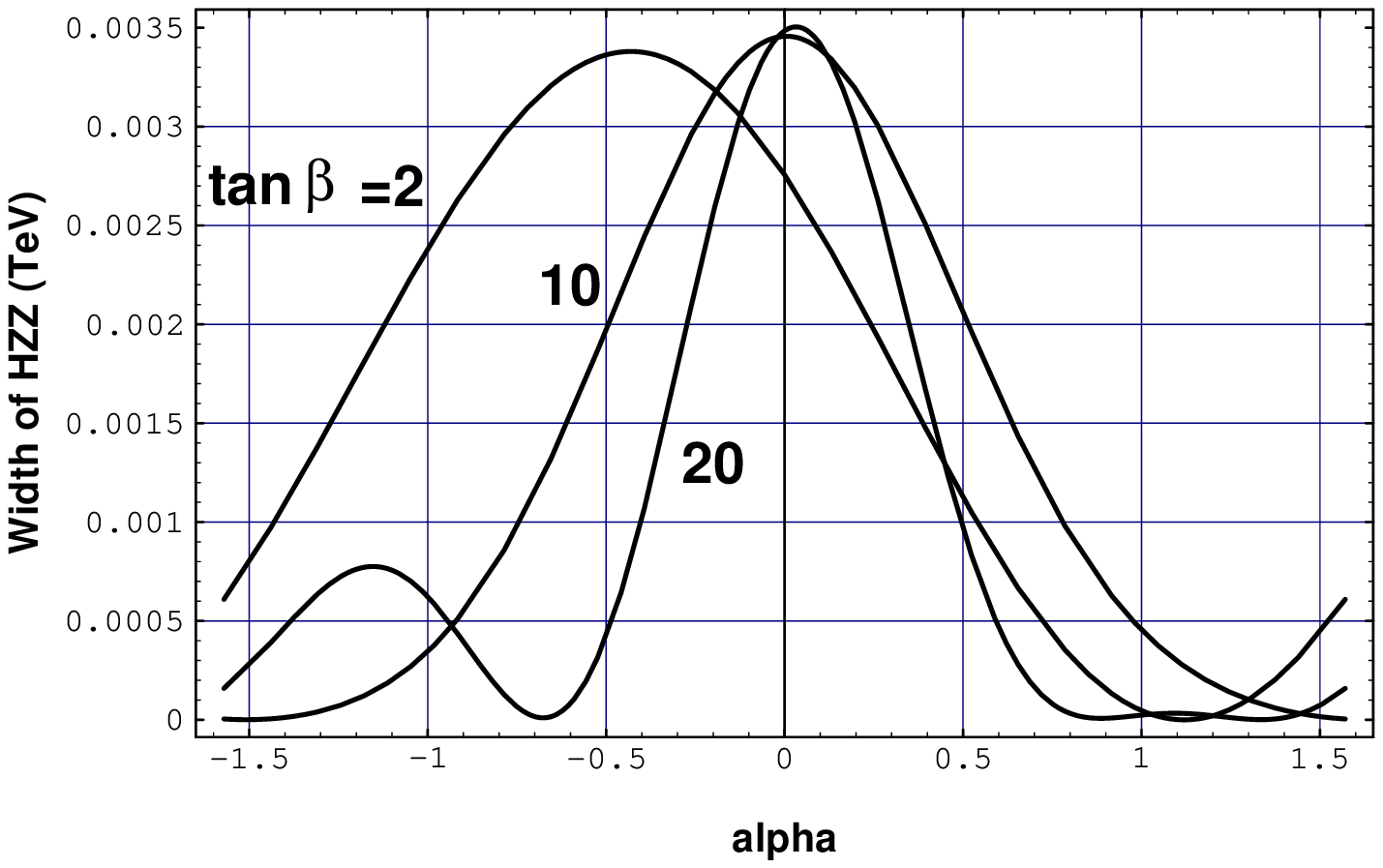}
\small
  {\bf Figure 2.}
The decay width $\Gamma (H \rightarrow Z_LZ_L)$ 
as a function of the mixing angle 
$\al$. The value of $\tan \be$ is fixed as (a) $\tan \be =2$, 
(b) $\tan \be =10$ and (c) $\tan \be =20$. 
The mass parameters are assumed as 
$m_H = 300$ GeV, $m_h = 400$ GeV, $m_G = 500$ GeV and $m_A = 600$ GeV.   
\normalsize
\end{minipage}
\end{center}

\section{Screening theorem for the vertices} 

\hspace{12pt}
The behavior of the decay widths $\Gamma (H \rightarrow W^+_LW^-_L)$ 
and $\Gamma (H \rightarrow Z_LZ_L)$  
will be dictated by the leading power contributions with respect to 
the masses of internal (heavy) 
scalars $G^{\pm}, A$ and $h$. 
We at first would like to discuss these contributions in the amplitudes 
${\cal M}_{Hww}$ and ${\cal M}_{Hzz}$,
which are of the form of $M^4/v^3$ (possibly times $\ln M$) 
on dimensional account  
($M$ represents collectively $m_G, m_A$ and/or $m_h$).   
These contributions are extracted from the full expression of the amplitudes 
as
\begin{eqnarray}
  {\cal M}_{Hww} (m_H^2) \longrightarrow & & 
   \fr{-1}{(4\pi)^2 v^3} \sn \ab \nn \\ 
  & & \hs \times \lt\{ (\ma - \mg) \ma -  \mg \ma \ln \fr{\ma}{\mg} \rt\}\nn\\
  & & + \;({\rm term\,\, from\,\, the\,\, prescription\,\, for\,\,}
            \delta \be ), 
\label{hml1} \\
{\cal M}_{Hzz} (m_H^2) \longrightarrow & &  0, \label{hml2}
\end{eqnarray} 
where the second term on RHS in \eq{hml1} 
has its origin form our prescription scheme for $\delta \beta$ 
by the $zA$-mixing. 
This term is extracted from the part 
which has the factor $\Pi_{zA}'(0) - \Pi_{wG}'(0)$,
where  $\Pi_{zA}(p^2)$ and $\Pi_{wG}(p^2)$ are two-point 
functions for $zA$ and $wG$ mixing respectively.

The leading contribution \eq{hml1} shows that there are mass-leading 
contributions in ${\cal M}_{Hww}$ except for the case of $m_G = m_A$.
Recall that $\Pi_{zA}(p^2)$ would equal $\Pi_{wG}(p^2)$ 
for $m_G = m_A$. 
This corresponds to the numerical results of 
$\Gamma (H \rightarrow W^+_LW^-_L)$, {\it i.e.}, 
radiative corrections are sensitive to $m_G$ and $m_A$ 
but insensitive to $m_h$
and are minimized in the case of $m_G = m_A$.   
On the other hand, \eq{hml2} 
shows that mass-leading 
contributions in ${\cal M}_{Hzz}$ always cancel out 
in accordance with the numerical results 
of $\Gamma (H \rightarrow Z_LZ_L)$, namely, radiative corrections 
are always small and insensitive to any of the masses of $G^{\pm}$, $A$ or  
$h$.

Now let us consider the reason for these cancellations of the leading-mass 
contributions in ${\cal M}_{Hww}(p^2)$ for $m_G = m_A$ and in 
${\cal M}_{Hzz}$ at any value of $m_G$ and $m_A$.  
Since the deviation from the mass degeneracy $m_G = m_A$ measures the 
custodial symmetry breaking, 
we may divide each amplitude into two parts as follows,
\begin{eqnarray}
  {\cal M}_{Hww} &=& {\cal M}^{\rm S} + {\cal M}_{Hww}^{\rm B}, 
\label{exp1}\\
  {\cal M}_{Hzz} &=& {\cal M}^{\rm S} + {\cal M}_{Hzz}^{\rm B}, 
\label{exp2}
\end{eqnarray}
where the superscript S means the custodial $SU(2)_V$ symmetric part and 
B stands for all the other (namely, $SU(2)_V$ breaking) part.   
Note that the first terms on RHS in eqs.\eq{exp1} and \eq{exp2} have 
to be equal to each other owing to the existence of the isospin  
symmetry between 
$w^{\pm}$ and $z$.
The leading contribution  \eq{hml1} 
necessarily comes from
${\cal M}^{\rm B}_{Hww}$. 
Both of \eq{exp1} and \eq{exp2} suggests that the mass-leading contributions 
to ${\cal M}^{\rm S}$ in eqs.\eq{exp1} and  \eq{exp2} always cancel out.
We are thus led to the new screening theorem for the vertices which we have 
presented in Introduction.

The mass-leading contributions seen in \eq{hml1} 
which have to belong to ${\cal M}_{Hww}^{\rm B}$  
must vanish in the custodial $SU(2)_V$ symmetric limit.
Thus the sensitivity to $m_G$ and $m_A$ and the correction-minimization 
for $m_G = m_A$ in $\Gamma (H \rightarrow W^+_LW^-_L)$ are both explained.   
As to $\Gamma (H \rightarrow Z_LZ_L)$,
the absence of leading-mass contributions, \eq{exp2}, which is valid even 
$m_G \neq m_A$, indicates that   the leading-mass contributions 
in ${\cal M}_{Hzz}^{\rm B}$ also cancel out in our scheme 
(on mass-shell, and definition of $\delta \be$ by the $zA$-mixing).  
This non-trivial cancellation in ${\cal M}_{Hzz}^{\rm B}$
can be proved on the one loop level by making use of 
renormalizability, absence  of the coupling $zw^+G^-$ and $zG^+G^-$ 
in the model and the screening theorem mentioned above. 
The proof will be presented in ref.\cite{kkt}.

Therefore, we have been able to 
explain all the characteristics of 
radiative corrections of the Higgs decay processes 
$H \rightarrow W^+_LW^-_L$ and $Z_LZ_L$ 
in terms of the screening theorem for vertices.

\section{Summary and Discussions}

\hspace{12pt}
In this talk, we have analyzed the radiative corrections to the decay processes
$H \rightarrow W^+_LW^-_L$ and $Z_LZ_L$ at one loop level in THDM and have 
investigated the sensitivity to the masses of the internal scalar bosons.
We have found that the radiative corrections to the decay width 
$\Gamma (H \rightarrow W^+_LW^-_L)$ are 
sensitive to the masses of $G^{\pm}$ and
$A$ but insensitive to the mass of $h$ 
and are minimized for $m_G = m_A$. 
On the other hand, it has also been found that 
the radiative corrections to 
$\Gamma (H \rightarrow Z_LZ_L)$ are insensitive to all the masses 
of internal scalar bosons and are always relatively small.

We have shown that these results are explained completely on the basis of  
the new screening theorem for vertices, which applies in  
the custodial $SU(2)_V$ symmetric limit of the model.    
This theorem for vertices, though we found it only at one loop level,  
is also likely to hold to all orders of the perturbation because of the 
similarity between this theorem and Veltman's screening theorem 
for oblique type corrections.

The one-loop insensitivity of $\Gamma (H \rightarrow Z_LZ_L)$ to 
the effect of the internal scalar bosons  in our calculation scheme 
could make this decay width to be a good experimental measuring tool 
for the mixing angle $\alpha$ (see fig.2). 
Then the experimental measurement of 
$\Gamma (H \rightarrow W^+_LW^-_L)$ could also provide us with 
 a good measure  for the custodial $SU(2)_V$ breaking.

\section*{\it Acknowledgments}  

\hspace{12pt}
The talker (S. K.) would like to thank M. Peskin for useful suggestions. 
This work is supported in part by the Grant in Aid for Scientific Research 
from the Ministry of Education, Science and Culture (Grant No.06640396).


\end{document}